\documentclass[10pt,a4paper]{article}
\usepackage[utf8]{inputenc}
\usepackage[english]{babel}
\usepackage{amsmath}
\usepackage{amsfonts}
\usepackage{amssymb}
\usepackage{amsthm}
\usepackage{placeins}
\usepackage{cite}
\usepackage{graphicx}
\usepackage{graphicx,overpic}
\usepackage{subfigure}
\usepackage{geometry}
\usepackage{color}
\usepackage{comment}

\title{\bf Nonlinear Lissajous orbits and particular superintegrability}
\author{A. M. Escobar-Ruiz$^{1}$, R. Azuaje$^{2}$}
\date{\small $^{1}$Departamento de F\'isica, Universidad Aut\'onoma Metropolitana--Iztapalapa, Mexico\\
admau@xanum.uam.mx\\[5pt]
$^{2}$Department of Physics, Faculty of Nuclear Sciences and Physical Engineering, Czech Technical University in Prague. Břehová 7, 115 19 Praha 1, Czech Republic\\
azuajraf@cvut.cz}

\begin{document}
\maketitle

\begin{abstract}
We investigate the geometry of classical trajectories generated by separable two-dimensional polynomial potentials of the form $V(x,y)=\tfrac{1}{2}\big(x^{2N}+A\,y^{2N}\big)$, where $N=1,2,\ldots,$ and $A>0$.
Special emphasis is placed on the emergence of nonlinear Lissajous figures and on the distinction between global and particular superintegrability.
In the harmonic case ($N=1$) closed periodic orbits are a consequence of an additional \emph{global} integral of motion whenever the frequency ratio is rational, rendering the system maximally superintegrable. In contrast, for anharmonic oscillators, already in the quartic case \((N=2)\), the oscillation frequencies depend on the partial energies, so periodic Lissajous-type trajectories occur only under nonlinear resonance conditions fixed by the initial data.
Accordingly, the extra conserved quantities that characterize these closed orbits are not global invariants but \emph{particular} (trajectory-dependent) integrals that emerge only on the resonant trajectories. For higher-degree potentials \(N\geq3\), the resonant trajectories are
naturally described by hyperelliptic phase constraints rather than by a
universal polynomial orbit equation.

\vspace{.3cm}
\end{abstract}

\section{Introduction}

Lissajous figures were introduced in 1857 by Jules Antoine Lissajous in his study of the composition of two mutually perpendicular harmonic vibrations; experimentally, such figures were made visible by attaching mirrors to perpendicular tuning forks and reflecting light from them \cite{lissajous1857memoire}. In classical mechanics, Lissajous figures arise as the configuration-space trajectories of a two-dimensional anisotropic harmonic oscillator \cite{Arnold78,GPS2002,landau1982mechanics,doll2007lissajous}. They remain central objects in several modern studies in classical and quantum mechanics \cite{deprit1991lissajous,krason2016generalization,escobar2021four,russo2023quantum,gallozzi2024between,turbiner2023quartic}. The distinction between rational and irrational frequency ratios already anticipates the resonance structure that plays a central role in the theory of nearly integrable Hamiltonian systems \cite{kolmogorov1954conservation,moser1962invariant,vladimir1963small}.

Lissajous figures constitute a paradigmatic example of ordered motion arising from the superposition of oscillations with commensurate frequencies. For the two-dimensional anisotropic harmonic oscillator, closure of the configuration-space orbit is a consequence of global superintegrability: whenever the frequency ratio \(\omega_x/\omega_y\) is rational, an additional integral of motion exists throughout phase space, and the bounded trajectories belong to closed Lissajous families determined by their amplitudes and relative phases \cite{rodriguez2008reduction,miller2013classical}. The closed trajectories are therefore not exceptional features of the dynamics, but systematic consequences of an underlying symmetry algebra \cite{daskaloyannis2007quantum,kalnins2005second,kress2007equivalence,marquette2010superintegrability,lopez2026tremblay,escobar2025two}. This global structure is well understood, and the harmonic case serves as the standard reference point for nonlinear generalizations.

Considerably less attention has been paid to nonlinear counterparts of Lissajous figures generated by anharmonic potentials. Anharmonic oscillators occupy a central place in nonlinear classical and quantum mechanics \cite{bender1969anharmonic,banerjee1978anharmonic,banerjee1978general,sergyeyev2007exact,turbiner2021anharmonic}, arising naturally in molecular physics, nonlinear optics, and perturbative treatments of field theories. In the anharmonic setting, the frequencies of oscillation are no longer fixed by the parameters of the potential alone, but depend on the energies of the corresponding one-dimensional motions. As a result, two trajectories with different partial energies generically oscillate with different frequency ratios, even when the potential is isotropic. The simple global geometric picture of harmonic Lissajous figures is therefore lost. Nevertheless, closed Lissajous-type trajectories may still persist, provided the initial data satisfy a nonlinear resonance condition. The natural questions are then: what geometric structure do these nonlinear closed orbits possess, and what conserved quantities, if any, are associated with them?

These questions touch on a fundamental issue in the theory of Hamiltonian systems: the relationship between the closure of trajectories and the existence of conserved quantities. For integrable systems in the Liouville sense, the existence of sufficiently many global integrals of motion constrains the dynamics to invariant tori. In two degrees of freedom, a rational frequency ratio selects resonant tori on which the motion is periodic. However, this standard action-angle description does not by itself imply the existence of an additional globally defined integral of motion. Superintegrable systems, namely systems possessing more independent integrals than degrees of freedom, exhibit stronger restrictions on their bounded trajectories \cite{miller2013classical}. The isotropic harmonic oscillator, and more generally the resonant anisotropic harmonic oscillator, belongs to this class. Once the potential departs from the harmonic form, however, additional global integrals are no longer expected in general. The persistence of closed orbits in such nonlinear systems therefore requires a more restricted explanation.

The notion of particular superintegrability provides such an explanation. A particular integral of motion is a phase-space function whose Poisson bracket with the Hamiltonian vanishes not identically throughout phase space, but only along specified trajectories or invariant submanifolds \cite{Turbiner2013}. While not a global symmetry, a particular integral encodes the fact that a selected part of the dynamics possesses additional conservation laws that are invisible on the full phase space \cite{escobar2024particular}. The concept of particular superintegrability, in which the number of such trajectory-dependent integrals exceeds the Liouville bound on a given invariant manifold, has begun to receive systematic attention \cite{turbiner2020particular,turbiner2021superintegrability}, but its geometric content and its connection with nonlinear resonance remain to be fully developed.

In this study, we analyze the classical dynamics of systems described by the Hamiltonian
\begin{equation}
H(x,y,p_x,p_y)=\frac{1}{2\,m}\left(p_x^2+p_y^2\right)+\frac{1}{2}\left(x^{2N}+A\,y^{2N}\right),
\label{eq:generalH}
\end{equation}
where \(N=1,2,\ldots\) and \(A>0\). We focus on the emergence of nonlinear Lissajous orbits and on the structure of the associated \textit{conserved quantities}. The separability of the potential implies the existence of two global partial-energy integrals in all cases. Beyond these integrals, however, the structure of conserved quantities depends critically on \(N\).

We compare three regimes. For \(N=1\), the system is the anisotropic harmonic oscillator, which is globally superintegrable whenever the frequency ratio is rational. For \(N=2\), the quartic anisotropic anharmonic oscillator \cite{kleinert1997resummation} is already non-isochronous: the frequency ratio depends on the partial energies, and closed nonlinear Lissajous curves occur only on resonant subsets of phase space. In this case, superintegrability becomes strictly trajectory-dependent, in the sense studied in the framework of particular (super)integrability, where only a part of the dynamics satisfies the conditions for complete integrability or superintegrability \cite{escobar2024particular}. The corresponding conserved quantities are particular integrals \cite{Turbiner2013}. For \(N\geq 3\), the same resonance mechanism persists, although the one-dimensional motions are described by hyperelliptic phase variables rather than elliptic ones.

Our main contributions are threefold. First, we derive the nonlinear resonance conditions for the separable polynomial potentials \eqref{eq:generalH}. These conditions depend not only on the anisotropy parameter \(A\), but also on the partial energies fixed by the initial data. We then construct phase-space quantities whose conservation holds precisely on the trajectories satisfying the nonlinear resonance conditions. Such quantities are particular integrals of motion.

Second, we describe the corresponding closed configuration-space resonant orbits. In the quartic case, where the dynamics is elliptic, the resonant orbits can be written explicitly as algebraic curves by using Jacobi elliptic multiplication formulas. For \(N\geq 3\), the motion is governed by hyperelliptic phase variables, and the resonant orbits are described by implicit phase constraints rather than by a universal polynomial equation \(F(x,y)=0\).

Third, we clarify why such trajectory-dependent invariants cannot, in general, be promoted to global conserved quantities. Taken together, these results show how integrability reorganizes itself in nonlinear separable systems: global superintegrability is lost, but closed resonant trajectories persist through particular integrals, namely quantities whose conservation is restricted to the imposition of the resonant condition.

The paper is organized as follows. Section~\ref{sec2} reviews the anisotropic harmonic oscillator \((N=1)\), emphasizing its global superintegrability for rational frequency ratios and the corresponding Lissajous orbit structure. Section~\ref{sec:quartic} analyzes the quartic oscillator \((N=2)\), where the motion is elliptic, the frequency ratio depends on the partial energies, and closed nonlinear Lissajous curves arise only on resonant orbits. We construct the corresponding particular integral of motion and give explicit algebraic forms for the resonant orbits on which such an integral is conserved. Section~\ref{sec:general} extends the analysis to \(N\geq 3\), where the one-dimensional motions are described by hyperelliptic phase variables and the resonant orbits are characterized by implicit phase constraints rather than by a universal polynomial equation. It also examines the restriction of these constructions to resonant trajectories and explains why the resulting trajectory-dependent relations cannot, in general, be promoted to global conserved quantities. Finally, Section~\ref{secconclusion} summarizes the conclusions.

\section{Case $N=1$: 2D anisotropic harmonic oscillator}
\label{sec2}

We consider the anisotropic harmonic oscillator with Hamiltonian
\begin{equation}
{\cal H}^{(N=1)}=\frac{1}{2m}\left(p_x^2+p_y^2\right)+\frac{1}{2}\left(x^2+A y^2\right),
\label{eq:Hharm}
\end{equation}
where $m>0$ is the particle mass, $A>0$ controls the anisotropy, and $(x,y;p_x,p_y)$ are canonical phase-space variables.
Hamilton's equations yield
\begin{equation}
\ddot{x}+\omega_x^2 x=0,\qquad \ddot{y}+\omega_y^2 y=0,
\label{eq:EOMharm}
\end{equation}
with natural frequencies
\begin{equation}
\omega_x=\sqrt{\frac{1}{m}},\qquad \omega_y=\sqrt{\frac{A}{m}}.
\label{eq:freqharm}
\end{equation}
For initial data at $t=0$ given by
\begin{equation}
x(0)=x_0,\quad p_x(0)=p_{x0},\qquad
y(0)=y_0,\quad p_y(0)=p_{y0},
\label{eq:IC}
\end{equation}
the solutions are
\begin{align}
x(t)&=x_0\cos(\omega_x t)+\frac{p_{x0}}{m\omega_x}\sin(\omega_x t),\label{eq:xt_harm}\\
y(t)&=y_0\cos(\omega_y t)+\frac{p_{y0}}{m\omega_y}\sin(\omega_y t).\label{eq:yt_harm}
\end{align}
The geometry of the configuration-space orbit is controlled by the frequency ratio
\begin{equation}
\frac{\omega_x}{\omega_y}=\frac{1}{\sqrt{A}}.
\label{eq:ratioharm}
\end{equation}
If \(\omega_x/\omega_y=p/q\), with coprime integers
\(p,q\in\mathbb{Z}_{>0}\), the motion is periodic with period 
\(T=2\pi p/\omega_x=2\pi q/\omega_y\), and the orbit closes to form a
Lissajous figure. If \(\omega_x/\omega_y\notin\mathbb{Q}\), the motion is
quasiperiodic on the invariant two-torus and its projection densely fills the
allowed rectangle \([-R_x,R_x]\times[-R_y,R_y]\) in configuration space (see below).

\subsection{Global integrals of motion and orbit equations}

\begin{figure}[h]
\centering
\includegraphics[width=0.9\textwidth]{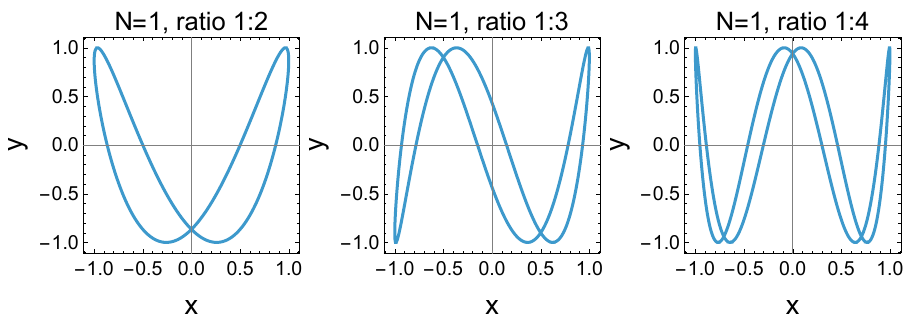}
\caption{
Closed configuration--space orbits of the anisotropic harmonic oscillator
(\(N=1\)) for rational frequency ratios \(\omega_x/\omega_y = 1{:}n\).
The anisotropy parameter is fixed to \(A=n^2\), ensuring commensurate
frequencies.
The trajectories are obtained from the exact analytic solutions
\(
x(t)=x_0\cos t+p_{x0}\sin t
\),
\(
y(t)=y_0\cos (nt)+\tfrac{p_{y0}}{n}\sin (nt)
\),
with initial conditions chosen so that the partial energies
\(
H_x=\tfrac12(p_x^2+x^2)
\)
and
\(
H_y=\tfrac12(p_y^2+A y^2)
\)
are conserved.
Panels correspond to (left) \(1{:}2\), (center) \(1{:}3\), and (right)
\(1{:}4\) resonances. Here \(m=1\) is used in the plotted trajectories.
}
\label{fig:harmonic-lissajous}
\end{figure}

Due to separability, the system admits the two (commuting) integrals
\begin{equation}
{\cal H}^{(N=1)}_x=\frac{p_x^2}{2m}+\frac{x^2}{2},\qquad
{\cal H}^{(N=1)}_y=\frac{p_y^2}{2m}+\frac{A y^2}{2},
\label{eq:partialE_harm}
\end{equation}
with ${\cal H}^{(N=1)}={\cal H}^{(N=1)}_x+{\cal H}^{(N=1)}_y$.
For resonant values of $A$ (equivalently $\omega_x/\omega_y\in\mathbb{Q}$), there exists a third \emph{global} integral $I_3$ that is conserved throughout phase space and makes the system maximally superintegrable. Representative closed harmonic Lissajous orbits for the resonances \(1:2\), \(1:3\), and \(1:4\) are shown in Fig.~\ref{fig:harmonic-lissajous}.

\paragraph{Orbit equation $y=y(x)$.}
For a given initial data~\eqref{eq:IC}, Eqs.~\eqref{eq:xt_harm}-\eqref{eq:yt_harm} can be written in amplitude-phase form
\begin{equation}
x(t)=R_x\cos(\omega_x t-\delta_x),\qquad
y(t)=R_y\cos(\omega_y t-\delta_y),
\label{eq:amp_phase}
\end{equation}
where the amplitudes and phases are fixed by the initial conditions,
\begin{equation}
R_x=\sqrt{x_0^2+\left(\frac{p_{x0}}{m\omega_x}\right)^2},\qquad
\tan\delta_x=\frac{p_{x0}}{m\omega_x x_0},
\label{eq:Rx_dx}
\end{equation}
and similarly
\begin{equation}
R_y=\sqrt{y_0^2+\left(\frac{p_{y0}}{m\omega_y}\right)^2},\qquad
\tan\delta_y=\frac{p_{y0}}{m\omega_y y_0}.
\label{eq:Ry_dy}
\end{equation}
Eliminating \(t\) from~\eqref{eq:amp_phase} yields the implicit Lissajous
relation
\begin{equation}
\cos\!\left(q\,\arccos\frac{x}{R_x}\right)
=
\cos\!\left(
p\,\arccos\frac{y}{R_y}-\Delta
\right),
\qquad
\Delta=q\delta_x-p\delta_y ,
\label{eq:lissajous_orbit_general}
\end{equation}
valid when \(\omega_x/\omega_y=p/q\), with
\(p,q\in\mathbb Z_{>0}\) coprime. Equation~\eqref{eq:lissajous_orbit_general}
defines the closed configuration-space curve traced by the motion. The
relative phase \(\Delta\) labels different members of the same resonance
class. Only an overall time translation is dynamically irrelevant; different
values of \(\Delta\) generally give geometrically distinct Lissajous curves.

For the representative zero-relative-phase branch, \(\Delta=0\), and for the
resonant values \(A=1,4,9\), corresponding to frequency ratios
\(1\!:\!1\), \(1\!:\!2\), and \(1\!:\!3\), the configuration-space orbits are
algebraic curves
\begin{equation}
f_A(x,y)=0 .
\end{equation}
The first cases are
\begin{align}
f_{1}(x,y)
&=
R_y x-R_x y,
\qquad &(A=1),
\label{eq:orbit_A1}
\\[0.2cm]
f_{4}(x,y)
&=
2R_y x^{2}-R_x^{2}y-R_x^{2}R_y,
\qquad &(A=4),
\label{eq:orbit_A4}
\\[0.2cm]
f_{9}(x,y)
&=
4R_y x^{3}-3R_y R_x^{2}x-R_x^{3}y,
\qquad &(A=9).
\label{eq:orbit_A9}
\end{align}
Here \(R_x\) and \(R_y\) are the oscillation amplitudes fixed by the initial
conditions. Other values of \(\Delta\) produce different algebraic curves in
the same resonance class. Thus the displayed polynomials should be understood
as representative phase-fixed Lissajous orbits. The polynomial degree grows
with the resonance order, reflecting the increasing algebraic complexity of
the closed harmonic trajectories.

\paragraph{Third global integral $I_3$.}
For the resonant anisotropic oscillator, one can construct a third conserved quantity $I_3$, polynomial in the phase--space variables $(x,y,p_x,p_y)$, which is globally defined throughout phase space.
For the lowest resonances we obtain:
\begin{itemize}
\item \textbf{$A=1$ (1:1):} the angular momentum,
\begin{equation}
I_3=L_z=xp_y-yp_x.
\label{eq:I3_A1}
\end{equation}

\item \textbf{$A=4$ (1:2):} a cubic integral,
\begin{equation}
I_3=p_x^{2}p_y+4m\,x y\,p_x-m\,x^{2}p_y.
\label{eq:I3_A4}
\end{equation}

\item \textbf{$A=9$ (1:3):} a quartic integral,
\begin{equation}
I_3=p_x^{3}p_y+9m\,x y\,p_x^{2}-3m\,x^{2}p_xp_y-3m^{2}x^{3}y.
\label{eq:I3_A9}
\end{equation}
\end{itemize}
In each case one verifies $\{{\cal H}^{(N=1)},I_3\}=0$ identically, so $I_3$ is conserved for all trajectories.
Thus, for these resonant values of $A$ the system admits three functionally independent integrals of motion for two degrees of freedom and is therefore maximally superintegrable.

\section{Case $N=2$: 2D quartic anisotropic anharmonic oscillator}
\label{sec:quartic}

We now turn to the quartic ($N=2$) separable oscillator described by the Hamiltonian
\begin{equation}
{\cal H}^{(N=2)}=\frac{1}{2m}\left(p_x^2+p_y^2\right)+\frac{1}{2}\left(x^4+A y^4\right),
\label{eq:Hquartic}
\end{equation}
where $m>0$ is the particle mass, $A>0$ controls the anisotropy, and $(x,y;p_x,p_y)$ are canonical phase-space variables. The corresponding Hamilton's equations imply
\begin{equation}
\dot{x}=\frac{p_x}{m},\qquad \dot{p}_x=-2x^3,\qquad
\dot{y}=\frac{p_y}{m},\qquad \dot{p}_y=-2A y^3,
\label{eq:HamEOM_quartic}
\end{equation}
or, equivalently,
\begin{equation}
\ddot{x}+\frac{2}{m}x^3=0,\qquad
\ddot{y}+\frac{2A}{m}y^3=0.
\label{eq:EOM_quartic}
\end{equation}
We specify initial data at $t=0$ by
\begin{equation}
x(0)=x_0,\quad p_x(0)=p_{x0},\qquad
y(0)=y_0,\quad p_y(0)=p_{y0}\ .
\label{eq:IC_quartic}
\end{equation}

\subsection{General solutions}

The conserved quantities ${\cal H}^{(N=2)}_x=E_x$ and
${\cal H}^{(N=2)}_y=E_y$, with
\begin{equation}
{\cal H}^{(N=2)}={\cal H}^{(N=2)}_x+{\cal H}^{(N=2)}_y,
\end{equation}
yield the first-order equations
\begin{equation}
\dot{x}^{\,2}=\frac{2E_x-x^4}{m},\qquad
\dot{y}^{\,2}=\frac{2E_y-Ay^4}{m}.
\label{eq:firstorder_quartic}
\end{equation}
The turning points are
\begin{equation}
x_{\max}=(2E_x)^{1/4},\qquad
y_{\max}=\left(\frac{2E_y}{A}\right)^{1/4}.
\label{eq:turningpoints}
\end{equation}
The solutions may be written in terms of Jacobi elliptic functions as
\begin{align}
x(t)&=x_{\max}\,
\mathrm{cn}\!\left(\alpha_x t+\varphi_x,k\right),
\label{eq:x_quartic_sol}\\
y(t)&=y_{\max}\,
\mathrm{cn}\!\left(\alpha_y t+\varphi_y,k\right),
\label{eq:y_quartic_sol}
\end{align}
with fixed modulus
\begin{equation}
k=\frac{1}{\sqrt{2}}.
\label{eq:kvalue}
\end{equation}
Here $\varphi_x$ and $\varphi_y$ are determined by the initial data.
The coefficients of the elliptic arguments are
\begin{equation}
\alpha_x=\sqrt{\frac{2}{m}}\,x_{\max}
=\frac{2^{3/4}E_x^{1/4}}{\sqrt{m}},
\qquad
\alpha_y=\sqrt{\frac{2A}{m}}\,y_{\max}
=\frac{2^{3/4}(AE_y)^{1/4}}{\sqrt{m}}.
\label{eq:elliptic_argument_coefficients}
\end{equation}
Indeed, since
\begin{equation}
\frac{d^2}{du^2}
\mathrm{cn}\!\left(u,\frac{1}{\sqrt2}\right)
+
\mathrm{cn}^3\!\left(u,\frac{1}{\sqrt2}\right)
=0,
\end{equation}
Eqs.~\eqref{eq:x_quartic_sol}--\eqref{eq:y_quartic_sol} solve
\begin{equation}
\ddot{x}+\frac{2}{m}x^3=0,\qquad
\ddot{y}+\frac{2A}{m}y^3=0.
\end{equation}

The real period of $\mathrm{cn}(u,k)$ is $4K(k)$, where $K(k)$ is
the complete elliptic integral of the first kind. Hence the physical
angular frequencies are
\begin{equation}
\Omega_x=\frac{\pi}{2K(k)}\,\alpha_x,\qquad
\Omega_y=\frac{\pi}{2K(k)}\,\alpha_y.
\label{eq:physical_frequencies}
\end{equation}
Equivalently,
\begin{equation}
\Omega_x=C\,\frac{E_x^{1/4}}{\sqrt{m}},
\qquad
\Omega_y=C\,\frac{(AE_y)^{1/4}}{\sqrt{m}},
\qquad
C=\frac{\pi\,2^{-1/4}}{K(1/\sqrt2)}.
\label{eq:freq_scaling}
\end{equation}
Consequently, the nonlinear frequency ratio is
\begin{equation}
\frac{\Omega_x}{\Omega_y}
=
\left(\frac{E_x}{AE_y}\right)^{1/4}.
\label{eq:nonlinear_ratio}
\end{equation}

\begin{figure}[t]
\centering
\includegraphics[width=0.9\textwidth]{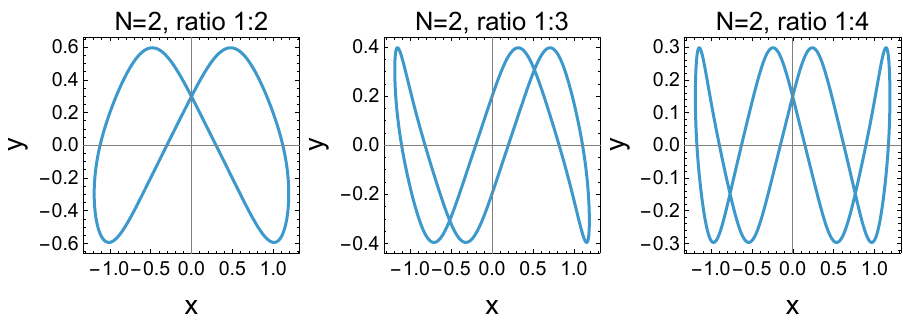}
\caption{
Closed configuration--space trajectories of the quartic anisotropic
oscillator (\(N=2\)) for nonlinear resonances
\(\Omega_x/\Omega_y = 1{:}n\).
The potential is
\(V(x,y)=\tfrac12(x^4 + A y^4)\), with equal partial energies
\(H_x = H_y = 1\).
The anisotropy parameter is fixed to \(A=n^4\), ensuring the nonlinear
commensurability condition.
Unlike the harmonic case, the frequencies depend on the energies, and
closed Lissajous--type orbits occur only for resonant initial data.
Panels correspond to (left) \(1{:}2\), (center) \(1{:}3\), and (right)
\(1{:}4\) resonances.
}
\label{fig:quartic-lissajous}
\end{figure}

\subsection{Particular integral of motion}
\label{subsec:particular}

For the quartic oscillator the globally defined Liouville integrals are the
two partial energies
\begin{equation}
{\cal H}^{(2)}_x
=
\frac{p_x^{2}}{2m}
+
\frac{x^{4}}{2},
\qquad
{\cal H}^{(2)}_y
=
\frac{p_y^{2}}{2m}
+
\frac{A y^{4}}{2},
\label{eq:Hxy_global}
\end{equation}
with
\begin{equation}
{\cal H}^{(2)}
=
{\cal H}^{(2)}_x
+
{\cal H}^{(2)}_y .
\end{equation}
For generic values of \(A\), no additional globally defined independent integral is expected in general. Nevertheless, one can
construct additional quantities whose Poisson bracket with the Hamiltonian vanishes only on resonant trajectories. Indeed, let us consider action-angle variables \((I_x,I_y,\theta_x,\theta_y)\) locally on a regular Liouville torus, away from the degenerate zero-partial-energy cases. We have
\begin{equation}
\dot{\theta}_{x}=\Omega_{x}(I_{x}), \quad \dot{\theta}_{y}=\Omega_{y}(I_{y}).
\end{equation}
We define the (possibly multi-valued \cite{tsiganov2008addition,tsiganov2009leonard}) function 
\begin{equation}
\label{Jparticular}
J^{(p,q)}=p\theta_{y}-q\theta_{x}, \ \ \textit{with}\ p,q\in\mathbb Z_{>0},\ \gcd(p,q)=1.
\end{equation}
The quantity $J^{(p,q)}$ is defined only modulo the period of the angle variables. Thus, it is a circle-valued resonant phase. Single-valued
representatives are given by
\begin{equation}
\cos J^{(p,q)},\qquad \sin J^{(p,q)}
\end{equation}
for $2\pi$-periodic action-angle variables, or by
\begin{equation}
\cos\left(\frac{\pi}{2K}J^{(p,q)}\right),\qquad
\sin\left(\frac{\pi}{2K}J^{(p,q)}\right)
\end{equation}
when the elliptic phases have a period $4K$.
We find that the single-valued representatives are particular integrals of motion. Indeed, for $J^{(p,q)}$ we have
\begin{equation}
\dot{J}^{(p,q)}=p\Omega_{y}-q\Omega_{x};
\end{equation}
which does not vanish all over the phase space; however, it does vanish for resonant trajectories, i.e., trajectories satisfying the nonlinear resonant condition 
\begin{equation}\label{eq:resonance_condition_quartic}
\frac{\Omega_x}{\Omega_y}=\frac{p}{q},\ \ p,q\in\mathbb Z_{>0},\ \gcd(p,q)=1.
\end{equation}

Representative closed quartic Lissajous-type trajectories satisfying this nonlinear resonance condition are shown in Fig.~\ref{fig:quartic-lissajous}.

\subsection{Resonant trajectories}

We now explicitly show the existence of resonant trajectories. In contrast with the harmonic case, the ratio~\eqref{eq:nonlinear_ratio} depends explicitly on the partial energies. Closed configuration-space orbits, or nonlinear Lissajous curves, occur only on trajectories satisfying the nonlinear resonance condition. We specialize to the equal-energy resonant family
\begin{equation}
E_x=E_y=1
\end{equation}
and to the resonant values
\begin{equation}
A=n^{4},
\qquad
n\in\mathbb{Z}_{>0}.
\label{eq:A_n4}
\end{equation}
Then
\begin{equation}
\frac{\Omega_x}{\Omega_y}
=
\frac{1}{n}.
\end{equation}
Thus \(A=1,2^4,3^4\) correspond respectively to the nonlinear resonances
\(1\!:\!1\), \(1\!:\!2\), and \(1\!:\!3\).

For \(E_x=E_y=1\) and \(A=n^4\), the turning points are
\begin{equation}
x_{\max}=2^{1/4},
\qquad
y_{\max}=\frac{2^{1/4}}{n}.
\end{equation}
It is convenient to introduce the dimensionless variables
\begin{equation}
\xi=\frac{x}{2^{1/4}},
\qquad
\eta=\frac{y}{y_{\max}}
=
\frac{n\,y}{2^{1/4}}.
\label{eq:dimensionless_quartic_variables}
\end{equation}
Using the elliptic solutions, the resonant trajectories may be written as
\begin{equation}
x(t)=x_{\max}\,\mathrm{cn}(u,k),
\qquad
y(t)=y_{\max}\,\mathrm{cn}(nu+\Delta,k),
\qquad
k=\frac{1}{\sqrt2},
\label{eq:quartic_cn_resonant_form}
\end{equation}
where
\begin{equation}
u=\alpha_x t+\varphi_x,
\qquad
\Delta=\varphi_y-n\varphi_x .
\end{equation}
The relative phase \(\Delta\) determines the particular member of the closed resonant family.

For the explicit orbit equations below, we take the phase-locked case
\begin{equation}
\Delta=0.
\end{equation}
Then
\begin{equation}
\xi=\mathrm{cn}(u,k),
\qquad
\eta=\mathrm{cn}(nu,k),
\qquad
k=\frac{1}{\sqrt2}.
\end{equation}
The multiplication theorem for Jacobi elliptic functions gives
\begin{equation}
\mathrm{cn}(nu,k)
=
R_n\!\left(\mathrm{cn}(u,k)\right),
\qquad
R_n(z)=\frac{P_n(z)}{Q_n(z)},
\label{eq:cn_multiplication_rational}
\end{equation}
where \(P_n\) and \(Q_n\) are polynomials depending on \(n\) and on the fixed modulus \(k=1/\sqrt2\). Therefore the phase-locked resonant orbit is defined implicitly by
\begin{equation}
F_{n^4}^{(0)}(x,y)
\equiv
\eta\,Q_n(\xi)-P_n(\xi)=0,
\label{eq:quartic_orbit_general_correct}
\end{equation}
with \(\xi\) and \(\eta\) given by \eqref{eq:dimensionless_quartic_variables}. The superscript \((0)\) indicates that the displayed equation corresponds to \(\Delta=0\). Other values of \(\Delta\) give different closed curves in the same resonance class; in general their explicit algebraic form is obtained by using the elliptic addition theorem before eliminating the phase.

For the first resonances one obtains the following explicit relations.

For \(A=1\), corresponding to the \(1\!:\!1\) resonance, one has
\begin{equation}
P_1(\xi)=\xi,
\qquad
Q_1(\xi)=1.
\end{equation}
Hence
\begin{equation}
F_1^{(0)}(x,y)=y-x=0.
\label{eq:quartic_orbit_1to1_correct}
\end{equation}

For \(A=2^4\), corresponding to the \(1\!:\!2\) resonance,
\begin{equation}
\mathrm{cn}(2u,k)
=
\frac{\xi^4+2\xi^2-1}{1+2\xi^2-\xi^4},
\qquad
k=\frac{1}{\sqrt2}.
\end{equation}
Therefore
\begin{equation}
\eta\left(1+2\xi^2-\xi^4\right)
-
\left(\xi^4+2\xi^2-1\right)=0.
\label{eq:quartic_orbit_1to2_dimensionless}
\end{equation}
In the original variables this becomes
\begin{equation}
2^{3/4}y
\left(
2+2\sqrt2\,x^2-x^4
\right)
-
x^4
-
2\sqrt2\,x^2
+
2
=0.
\label{eq:quartic_orbit_1to2_correct}
\end{equation}

For \(A=3^4\), corresponding to the \(1\!:\!3\) resonance,
\begin{equation}
\mathrm{cn}(3u,k)
=
\frac{\xi\left(\xi^8+6\xi^4-3\right)}
{1+6\xi^4-3\xi^8},
\qquad
k=\frac{1}{\sqrt2}.
\end{equation}
Thus
\begin{equation}
\eta\left(1+6\xi^4-3\xi^8\right)
-
\xi\left(\xi^8+6\xi^4-3\right)=0.
\label{eq:quartic_orbit_1to3_dimensionless}
\end{equation}
In the original variables this gives
\begin{equation}
3y
\left(
4+12x^4-3x^8
\right)
-
x
\left(
x^8+12x^4-12
\right)
=0.
\label{eq:quartic_orbit_1to3_correct}
\end{equation}

Unlike the harmonic oscillator, the closure of these quartic trajectories is not determined by a fixed frequency ratio in the potential alone. It is selected by the nonlinear energy-dependent resonance condition \eqref{eq:resonance_condition_quartic}. Consequently, the closed curves are associated with resonant trajectories rather than with a global superintegrability structure.

\paragraph{Lowest resonances.}

We now rewrite the preceding phase-locked orbit constraints as phase-space representatives on the corresponding energy shells. These representatives are not global integrals. Their Poisson brackets with the Hamiltonian vanish only after restriction to the selected phase-locked resonant trajectory.

For \(n=1\), \(A=1\), corresponding to the \(1\!:\!1\) resonance, the phase-locked orbit is
\begin{equation}
F^{(0)}_{1}(x,y)=y-x=0.
\label{eq:F1_particular}
\end{equation}
Equivalently, this branch is the invariant diagonal submanifold
\begin{equation}
x=y,
\qquad
p_x=p_y.
\end{equation}
On this submanifold the angular momentum
\begin{equation}
L_z=xp_y-yp_x
\label{eq:Lz_particular}
\end{equation}
vanishes and is conserved. In the full phase space, however,
\begin{equation}
\left\{
{\cal H}^{(2)}_{A=1},L_z
\right\}
=
2xy\left(y^2-x^2\right),
\label{eq:H_Lz_quartic_correct}
\end{equation}
which is not identically zero. Thus \(L_z\) is not a global integral of the quartic oscillator, although it becomes conserved on the invariant diagonal submanifolds
\begin{equation}
x=\pm y,\qquad p_x=\pm p_y .
\end{equation}

For \(n=2\), \(A=2^4\), corresponding to the \(1\!:\!2\) resonance, the phase-locked orbit equation is
\begin{equation}
F^{(0)}_{2^4}(x,y)
=
2^{3/4}y
\left(
2+2\sqrt2\,x^2-x^4
\right)
-
x^4
-
2\sqrt2\,x^2
+
2
=
0.
\label{eq:F2_particular_coordinate}
\end{equation}
Using the energy-shell identity
\begin{equation}
x^4=2-\frac{p_x^2}{m},
\end{equation}
an equivalent phase-space representative on the branch \(E_x=1\) is
\begin{equation}
\mathcal I^{(0)}_{1:2}(x,y,p_x)
=
2^{3/4}y
\left(
\frac{p_x^2}{m}
+
2\sqrt2\,x^2
\right)
+
\frac{p_x^2}{m}
-
2\sqrt2\,x^2 .
\label{eq:I12_particular}
\end{equation}
It satisfies
\begin{equation}
\left.
\left\{
{\cal H}^{(2)}_{A=2^4},\mathcal I^{(0)}_{1:2}
\right\}
\right|_{\gamma_{1:2}^{(0)}}
=
0,
\end{equation}
where \(\gamma_{1:2}^{(0)}\) denotes the corresponding phase-locked resonant trajectory. The Poisson bracket is not zero as a global identity on phase space.

For \(n=3\), \(A=3^4\), corresponding to the \(1\!:\!3\) resonance, the phase-locked orbit equation is
\begin{equation}
F^{(0)}_{3^4}(x,y)
=
3y
\left(
4+12x^4-3x^8
\right)
-
x
\left(
x^8+12x^4-12
\right)
=
0.
\label{eq:F3_particular_coordinate}
\end{equation}
Using again
\begin{equation}
x^4=2-\frac{p_x^2}{m},
\end{equation}
this may be written on the energy shell as
\begin{equation}
\mathcal I^{(0)}_{1:3}(x,y,p_x)
=
3y
\left[
16
-
3\left(\frac{p_x^2}{m}\right)^2
\right]
-
x
\left[
16
-
16\frac{p_x^2}{m}
+
\left(\frac{p_x^2}{m}\right)^2
\right].
\label{eq:I13_particular}
\end{equation}
Again,
\begin{equation}
\left.
\left\{
{\cal H}^{(2)}_{A=3^4},\mathcal I^{(0)}_{1:3}
\right\}
\right|_{\gamma_{1:3}^{(0)}}
=
0,
\end{equation}
while the bracket does not vanish on the full phase space.

For higher resonances the construction is analogous. One first obtains \(P_n\) and \(Q_n\) from the \(\mathrm{cn}\)-multiplication formula at \(k=1/\sqrt2\), forms the phase-locked orbit constraint
\[
F^{(0)}_{n^4}(x,y)=\eta Q_n(\xi)-P_n(\xi),
\]
and, if desired, uses the energy-shell identities
\[
x^4=2-\frac{p_x^2}{m},
\qquad
A y^4=2-\frac{p_y^2}{m},
\]
to obtain a phase-space representative. These functions are conserved only on the corresponding phase-locked resonant trajectories. They are therefore particular, not global, integrals of motion.

\section{General case $N\geq 3$}
\label{sec:general}

Let us consider the general Hamiltonian function
\begin{equation}
{\cal H}^{(N)}=\frac{1}{2m}\left(p_x^2+p_y^2\right)
+\frac{1}{2}\left(x^{2N}+A y^{2N}\right),
\label{eq:H_gen}
\end{equation}
for $N\geq 3$, where $A>0$ is the anisotropy parameter.
As in the previous cases, separability implies the existence of the two global
integrals
\begin{equation}
{\cal H}^{(N)}_x=\frac{p_x^2}{2m}+\frac{x^{2N}}{2},\qquad
{\cal H}^{(N)}_y=\frac{p_y^2}{2m}+A\frac{y^{2N}}{2},
\label{eq:Hxy_gen}
\end{equation}
with ${\cal H}^{(N)}={\cal H}^{(N)}_x+{\cal H}^{(N)}_y$.

\subsection{Equations of motion and first integrals}

Hamilton’s equations yield
\begin{equation}
\ddot{x}+\frac{N}{m}x^{2N-1}=0,\qquad
\ddot{y}+A\frac{N}{m}y^{2N-1}=0.
\label{eq:EOM_gen}
\end{equation}
Using the conserved partial energies, the dynamics may be written in first--order
form as
\begin{equation}
\dot{x}^{2}=\frac{2}{m}E_x-\frac{1}{m}x^{2N},\qquad
\dot{y}^{2}=\frac{2}{m}E_y-\frac{A}{m}y^{2N}.
\label{eq:firstorder_gen}
\end{equation}
The turning points are
\begin{equation}
x_{\max}=(2E_x)^{1/(2N)},\qquad
y_{\max}=\left(\frac{2E_y}{A}\right)^{1/(2N)}.
\label{eq:turning_gen}
\end{equation}

\subsection{Frequency scaling and nonlinear resonance}

Systems with $N\geq 3$ are non--isochronous. From
Eq.~\eqref{eq:firstorder_gen}, the period of the $x$--motion is
\begin{equation}
T_x
=
4\sqrt{m}
\int_{0}^{(2E_x)^{1/(2N)}}
\frac{dx}{\sqrt{2E_x-x^{2N}}}.
\end{equation}
After the change of variables
\[
x=(2E_x)^{1/(2N)}u,
\]
one obtains
\begin{equation}
T_x
=
4\sqrt{m}\,
(2E_x)^{\frac{1}{2N}-\frac{1}{2}}
\int_0^1
\frac{du}{\sqrt{1-u^{2N}}}.
\end{equation}
Hence the characteristic angular frequency scales as
\begin{equation}
\Omega_x
=
C_N\,\frac{E_x^{(N-1)/(2N)}}{\sqrt{m}},
\label{eq:freq_x_gen_correct}
\end{equation}
where $C_N>0$ is a numerical constant depending only on $N$.

\begin{figure}[h]
\centering
\includegraphics[width=\textwidth]{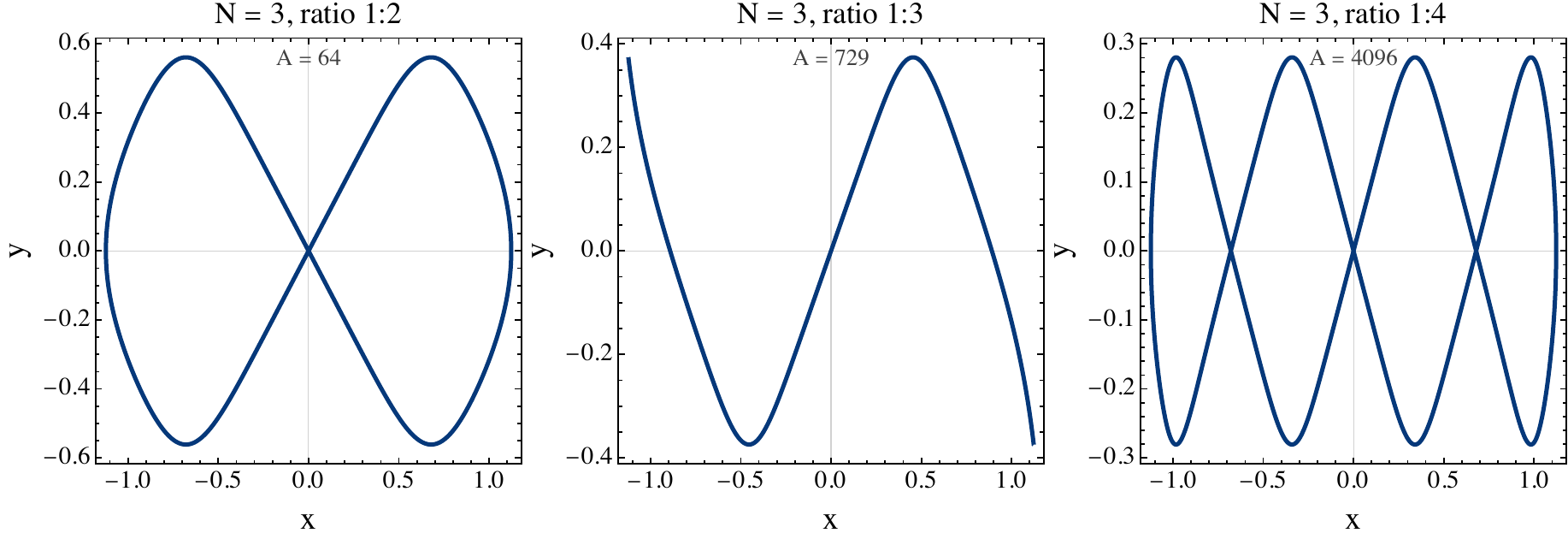}
\caption{
Closed configuration-space trajectories of the sextic anisotropic oscillator
\((N=3)\) for the resonances \(\Omega_x/\Omega_y=1{:}n\).
We set \(E_x=E_y=1\) and \(A=n^6\), corresponding to \(1{:}2\), \(1{:}3\),
and \(1{:}4\). These curves illustrate the hyperelliptic nonlinear Lissajous regime
associated with the phase relation~\eqref{eq:orbit_general_implicit}.
}
\label{fig:sextic_lissajous}
\end{figure}

Similarly, for the $y$--motion,
\begin{equation}
T_y
=
4\sqrt{m}
\int_{0}^{(2E_y/A)^{1/(2N)}}
\frac{dy}{\sqrt{2E_y-Ay^{2N}}}.
\end{equation}
Using
\[
y=\left(\frac{2E_y}{A}\right)^{1/(2N)}v,
\]
we find
\begin{equation}
T_y
=
4\sqrt{m}\,
A^{-1/(2N)}
(2E_y)^{\frac{1}{2N}-\frac{1}{2}}
\int_0^1
\frac{dv}{\sqrt{1-v^{2N}}}.
\end{equation}
Therefore
\begin{equation}
\Omega_y
=
C_N\,\frac{A^{1/(2N)}E_y^{(N-1)/(2N)}}{\sqrt{m}}.
\label{eq:freq_y_gen_correct}
\end{equation}
Consequently, the frequency ratio is
\begin{equation}
\frac{\Omega_x}{\Omega_y}
=
A^{-1/(2N)}
\left(\frac{E_x}{E_y}\right)^{(N-1)/(2N)}.
\label{eq:ratio_gen_correct}
\end{equation}
Closed configuration--space orbits occur only when this ratio is rational,
\begin{equation}
\frac{\Omega_x}{\Omega_y}
=
\frac{p}{q},
\qquad
p,q\in\mathbb{Z}_{>0},
\qquad
\gcd(p,q)=1.
\label{eq:resonance_gen_correct}
\end{equation}
Thus, unlike the harmonic oscillator, the resonance condition depends on the
partial energies as well as on the anisotropy parameter $A$. Examples of closed sextic nonlinear Lissajous trajectories for \(N=3\) are shown in Fig.~\ref{fig:sextic_lissajous}.

\begin{figure}[h]
\centering
\includegraphics[width=0.72\textwidth]{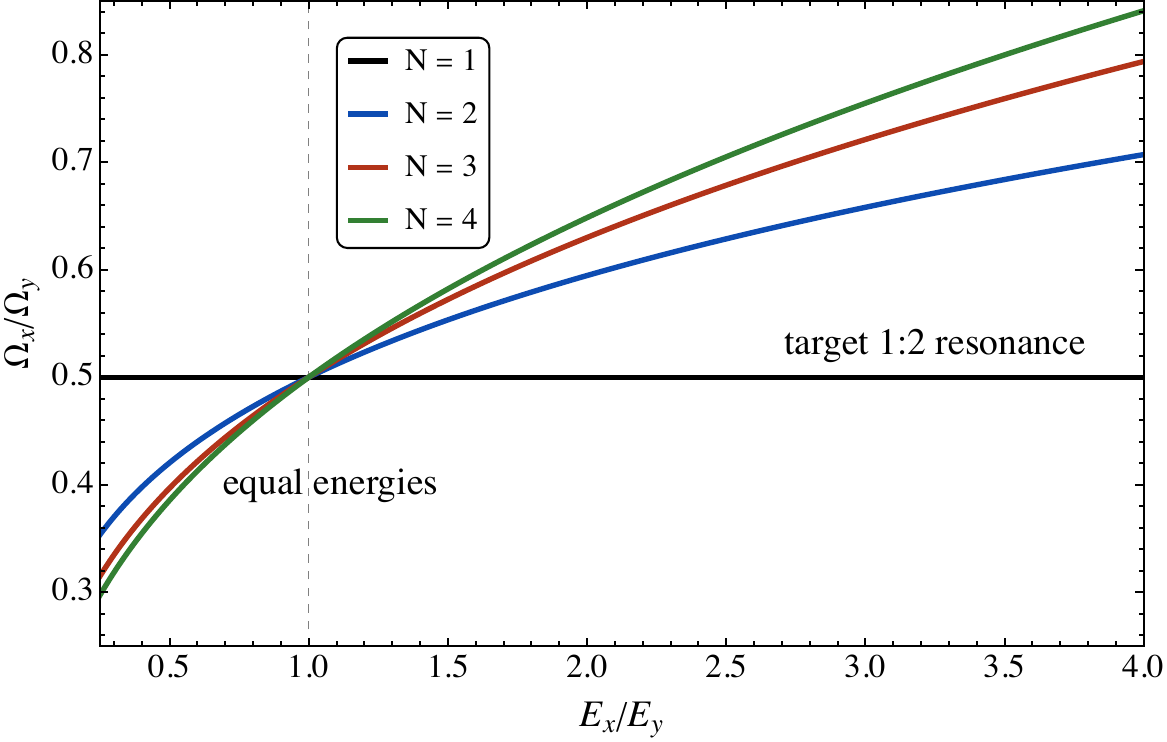}
\caption{
Energy dependence of the frequency ratio
\(\Omega_x/\Omega_y\) for the family
\(V(x,y)=\frac12(x^{2N}+Ay^{2N})\). For each \(N\), we choose
\(A=2^{2N}\), so that the equal-energy shell \(E_x=E_y\) realizes the
\(1{:}2\) resonance. The harmonic case \(N=1\) is energy independent,
whereas for \(N\geq2\) the resonance is selected by the partial energies.
}
\label{fig:frequency_scaling}
\end{figure}

\begin{figure}[h]
\centering
\includegraphics[width=1.0\textwidth]{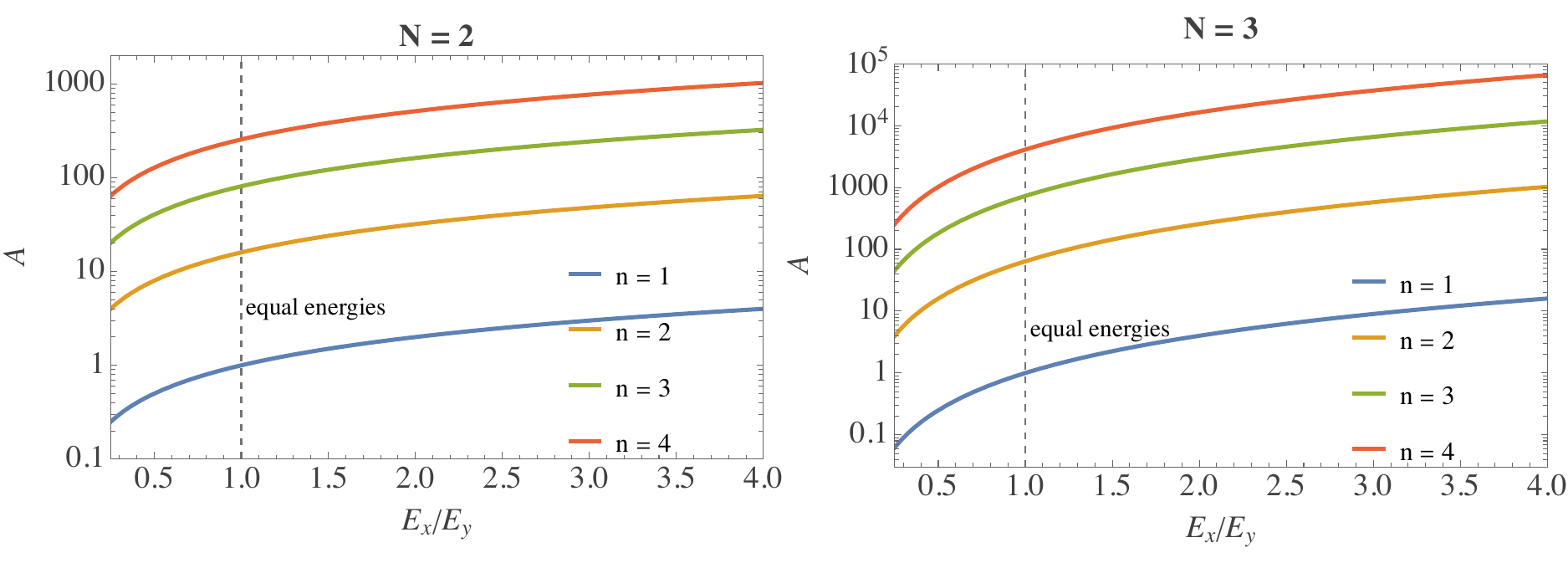}
\caption{
Resonance curves in the \((E_x/E_y,A)\) plane for the nonlinear oscillators
\(V(x,y)=\frac12(x^{2N}+Ay^{2N})\). The curves are defined by
$
A=n^{2N}\left(E_x/E_y\right)^{N-1},
$
which is equivalent to the nonlinear resonance condition
\(\Omega_x/\Omega_y=1/n\). The dashed vertical line marks the equal-energy
shell \(E_x=E_y\), and the marked points indicate the corresponding resonant
values \(A=n^{2N}\). The plot shows that nonlinear Lissajous trajectories are
selected jointly by the anisotropy parameter and the partial-energy ratio.
}
\label{fig:resonance_manifolds}
\end{figure}

Equivalently, for a prescribed \(1:n\) resonance, Eq.~\eqref{eq:ratio_gen_correct}
can be solved for the anisotropy parameter:
\begin{equation}
A
=
n^{2N}
\left(\frac{E_x}{E_y}\right)^{N-1}.
\label{eq:A_resonance_curve}
\end{equation}
This expression shows explicitly that the nonlinear resonance is selected
jointly by the anisotropy parameter and the partial-energy ratio. Figures~\ref{fig:frequency_scaling} and~\ref{fig:resonance_manifolds}
illustrate the energy dependence of the nonlinear resonance condition.
Figure~\ref{fig:frequency_scaling} shows the frequency ratio
\(\Omega_x/\Omega_y\) as a function of the partial-energy ratio \(E_x/E_y\)
for several values of \(N\), with the anisotropy chosen so that the
equal-energy shell realizes the \(1\!:\!2\) resonance. The harmonic case
\(N=1\) is independent of \(E_x/E_y\), whereas for \(N\geq2\) the frequency
ratio varies with the energy distribution. Equivalently,
Figure~\ref{fig:resonance_manifolds} displays the corresponding resonance
curves in the \((E_x/E_y,A)\) plane. These curves show that, in the nonlinear
systems, closed Lissajous-type trajectories are selected jointly by the
anisotropy parameter and the partial-energy ratio, rather than by the
potential parameter \(A\) alone.

\FloatBarrier

\subsection{Particular integral and resonant trajectories}

We now fix equal partial energies,
\begin{equation}
E_x=E_y=1,
\label{eq:energies_gen}
\end{equation}
and impose a \(1\!:\!n\) nonlinear resonance,
\begin{equation}
\frac{\Omega_x}{\Omega_y}
=
\frac{1}{n},
\qquad
n\in\mathbb{Z}_{>0}.
\label{eq:resonance_gen}
\end{equation}
Using Eq.~\eqref{eq:A_resonance_curve} on the equal-energy shell
\(E_x=E_y=1\), one obtains
\begin{equation}
A=n^{2N}.
\label{eq:A_ngen}
\end{equation}
For any \(n\in\mathbb{N}\), we shall denote
\begin{equation}
A_n=n^{2N}.
\end{equation}

For \(E_x=E_y=1\) and \(A=A_n\), the turning points are
\begin{equation}
x_{\max}=2^{1/(2N)},
\qquad
y_{\max}
=
\left(\frac{2}{A_n}\right)^{1/(2N)}
=
\frac{2^{1/(2N)}}{n}.
\label{eq:turningpoints_gen_res}
\end{equation}
We introduce the dimensionless variables
\begin{equation}
\xi=\frac{x}{2^{1/(2N)}},
\qquad
\eta=\frac{y}{y_{\max}}
=
\frac{n\,y}{2^{1/(2N)}}.
\label{eq:dimensionless_gen}
\end{equation}

For \(N\geq 3\), the separated one-dimensional motions are described by
hyperelliptic integrals. Let
\begin{equation}
K_N
=
\int_0^1
\frac{ds}{\sqrt{1-s^{2N}}},
\label{eq:KN_def}
\end{equation}
and define the generalized cosine function \(\mathrm{C}_N(u)\) as the
real \(4K_N\)-periodic solution of
\begin{equation}
\left(\frac{d\,\mathrm{C}_N}{du}\right)^2
=
1-\mathrm{C}_N^{2N},
\qquad
\mathrm{C}_N(0)=1,
\qquad
\mathrm{C}_N'(0)=0.
\label{eq:generalized_cosine_def}
\end{equation}
Equivalently, on the monotonic branch \(0\leq u\leq K_N\),
\begin{equation}
u
=
\int_{\mathrm{C}_N(u)}^1
\frac{ds}{\sqrt{1-s^{2N}}}.
\label{eq:CN_inverse}
\end{equation}

With these definitions, the resonant solutions can be written as
\begin{equation}
x(t)
=
2^{1/(2N)}\,\mathrm{C}_N(u),
\qquad
y(t)
=
\frac{2^{1/(2N)}}{n}\,
\mathrm{C}_N(nu+\Delta),
\label{eq:resonant_solutions_gen}
\end{equation}
where
\begin{equation}
u=
\frac{2^{(N-1)/(2N)}}{\sqrt{m}}\,t+\varphi_x,
\qquad
\Delta=\varphi_y-n\varphi_x.
\end{equation}
The constant \(\Delta\) is the relative phase and determines the particular
member of the closed resonant family.

Eliminating the time variable gives the implicit orbit relation
\begin{equation}
\eta
=
\mathrm{C}_N
\left(
n\,\mathrm{C}_N^{-1}(\xi)+\Delta
\right),
\label{eq:orbit_general_implicit}
\end{equation}
with the appropriate branch of \(\mathrm{C}_N^{-1}\) chosen along each
monotonic segment of the motion. Equivalently, on a monotonic branch,
\begin{equation}
\int_{\eta}^{1}
\frac{ds}{\sqrt{1-s^{2N}}}
=
n
\int_{\xi}^{1}
\frac{ds}{\sqrt{1-s^{2N}}}
+\Delta
\qquad
\mathrm{mod}\;4K_N.
\label{eq:orbit_general_integral}
\end{equation}
Equations~\eqref{eq:orbit_general_implicit} and
\eqref{eq:orbit_general_integral} define the configuration-space orbit
\begin{equation}
F_n^N(x,y;\Delta)=0.
\label{eq:orbit_general}
\end{equation}

Thus, for \(N\geq 3\), the orbit equation is naturally expressed in terms of
hyperelliptic phase variables. In special low-order cases the implicit
relation may reduce to an algebraic equation, but there is no universal
polynomial formula of the form
\[
y^2(2+x^{2N})^n
-
C_n^N x^2(2-x^{2N})^{n-1} 
=0 \ ,
\]
for the general \(N\geq 3\) system. The mass \(m\) fixes only the time scale
and drops out of the orbit geometry.

The resonant phase integral is
\[
J^{(N,n)}=\theta_y-n\theta_x \mod 4K_N .
\]
Since, on the resonant shell,
\[
\dot\theta_y=n\dot\theta_x,
\]
we have
\[
\dot J^{(N,n)}=0.
\]
For a fixed relative phase \(\Delta\), the phase-locked component is defined by
\[
J^{(N,n)}=\Delta \mod 4K_N.
\]
Single-valued representatives are
\[
I^{(N,n)}_{\Delta,c}
=
\cos\left[
\frac{\pi}{2K_N}
(\theta_y-n\theta_x-\Delta)
\right],
\]
and
\[
I^{(N,n)}_{\Delta,s}
=
\sin\left[
\frac{\pi}{2K_N}
(\theta_y-n\theta_x-\Delta)
\right].
\]
Locally on each phase branch, these functions satisfy
\[
\{H^{(N)},I^{(N,n)}_{\Delta,c}\}=0,
\qquad
\{H^{(N)},I^{(N,n)}_{\Delta,s}\}=0
\]
after restriction to the resonant shell.

They are not conserved as global functions on the full phase space. They are therefore particular integrals associated with the resonant condition.

On a fixed monotonic branch, this construction can be written directly in
terms of hyperelliptic integrals. Define
\begin{equation}
{\cal A}_N(z)
=
\int_z^1
\frac{ds}{\sqrt{1-s^{2N}}}.
\label{eq:AN_def}
\end{equation}
Then the phase constraint becomes
\begin{equation}
{\cal A}_N(\eta)
-
n\,{\cal A}_N(\xi)
-
\Delta
=
0
\qquad
\mathrm{mod}\;4K_N.
\label{eq:integral_particular_general}
\end{equation}
Equation~\eqref{eq:integral_particular_general} is the hyperelliptic analogue
of the algebraic orbit constraint used in the harmonic and quartic cases.

\paragraph{Lowest resonances.}

For \(n=1\), one has \(A=1\). The particular phase integral is
\begin{equation}
J^{(N,1)}=\theta_y-\theta_x
\qquad
\mathrm{mod}\;4K_N.
\end{equation}
For the phase-locked choices \(\Delta=0\) and \(\Delta=2K_N\), this includes the invariant diagonal submanifolds
\[
x=y,\qquad p_x=p_y,
\]
and
\[
x=-y,\qquad p_x=-p_y,
\]
respectively. Along these invariant
lines the angular momentum
\begin{equation}
L_z=xp_y-yp_x \ ,
\end{equation}
vanishes and remains constant. In the full phase space, however,
\begin{equation}
\left\{
{\cal H}^{(N)},L_z 
\right\}
=
Nxy\left(y^{2N-2}-x^{2N-2}\right),
\label{eq:HLz_general_correct}
\end{equation}
which is not identically zero for \(N\geq2\). Thus \(L_z\) is not a global
integral of the nonlinear oscillator, although it becomes a particular
integral on the special invariant lines.

For \(n=2\), the resonant anisotropy is
\begin{equation}
A=2^{2N},
\end{equation}
and the corresponding particular integral is
\begin{equation}
J^{(N,2)}=\theta_y-2\theta_x
\qquad
\mathrm{mod}\;4K_N.
\end{equation}
For the sextic oscillator \(N=3\), this gives \(A=2^6=64\).

For \(n=3\), the resonant anisotropy is
\begin{equation}
A=3^{2N},
\end{equation}
and the corresponding particular integral is
\begin{equation}
J^{(N,3)}=\theta_y-3\theta_x
\qquad
\mathrm{mod}\;4K_N.
\end{equation}
For the sextic oscillator \(N=3\), this gives \(A=3^6=729\).

Thus, for \(N\geq3\), the particular integrals are naturally expressed as
hyperelliptic phase constraints rather than as universal polynomial functions
of \(x\) and \(y\). The mass \(m\) fixes only the common time scale and does
not enter the orbit geometry or the resonant phase constraint.

\subsection{Interpretation}

The preceding examples show that, after imposing a resonant condition, additional trajectory-dependent relations may appear among the phase-space variables. In the quartic case these relations can be written as algebraic orbit constraints, whereas for \(N\geq 3\) they are naturally expressed as hyperelliptic phase constraints.

These relations reflect the loss of functional independence produced by the restriction to a lower-dimensional invariant set. They do not represent additional global symmetries, nor do they imply that the nonlinear system inherits the global superintegrability of the harmonic oscillator.

It is important to distinguish the construction of the presented particular integrals from the tautological fact that a function vanishing on a single trajectory is constant along that trajectory. The particular integrals considered here arise from the phase variables of the separated motions and from the nonlinear resonance condition. On a resonant set the phases satisfy \(\dot{\theta}_y=n\dot{\theta}_x\), so the combination \(J^{(N,n)}=\theta_y-n\theta_x\) is conserved modulo the corresponding period. Thus, the invariant is generated by phase locking on a resonant set, not by an arbitrary defining equation for an individual trajectory.

In this sense, particular integrability in nonlinear separable systems is a trajectory-level phenomenon generated by nonlinear resonance. The resonance condition selects special resonant trajectories, and the corresponding particular integrals characterize those trajectories rather than providing global labels for all trajectories.

\section{Discussion and Conclusions}
\label{secconclusion}

We have shown that nonlinear Lissajous orbits arise naturally in separable anharmonic systems through nonlinear resonance. In the harmonic oscillator, closed trajectories follow from global superintegrability: additional integrals exist on the full phase space and organize complete families of periodic motions. In anharmonic systems, the frequencies depend on the partial energies, so closure occurs only on resonant invariant trajectories.

The main result is the identification of how the superintegrable structure is reorganized once global superintegrability is lost. For \(N=1\), rational frequency ratios give global additional integrals and hence closed Lissajous trajectories throughout the bounded region of phase space. For \(N\geq 2\), the system remains completely Liouville integrable by separability, but the additional structures associated with closed motion become restricted to resonant sets.

In the quartic case, these resonant sets can be described by algebraic constraints obtained from elliptic multiplication formulas. For \(N\geq 3\), the corresponding constraints are naturally written in terms of hyperelliptic phase variables. In both cases, the associated particular integrals are not global conserved charges. Rather, they encode the restricted conservation laws that hold on the resonant trajectories supporting closed orbits.

From this viewpoint, the fact that a particular integral $\cal I$ may vanish on its associated trajectory is not a defect. Its content lies in the restricted identity
\[
\{{\cal I},H\}=0 \ ,
\]
on the selected resonant set, while \(\{{\cal I},H\}\) does not vanish identically on the full phase space. The particular integral, therefore, detects the closure of the dynamics on a lower-dimensional set.

Hence, global superintegrability is not preserved by the nonlinear deformation, but its trace survives on resonant sets. Global partial-energy integrals provide Liouville integrability, while trajectory-dependent particular integrals describe the exceptional closed orbits selected by nonlinear resonance. 

This work also opens several directions. A first one is the quantum counterpart of nonlinear resonant conditions and their particular integrals, where one may ask whether restricted classical invariants leave signatures in spectral degeneracies, scarred states, or quasi-exactly solvable sectors. A second direction is the extension beyond separable polynomial potentials, where resonant non-empty sets may persist only approximately but still organize long-lived coherent motion. More broadly, the results point to a classification problem for nonlinear Hamiltonian systems in which global superintegrability is lost, while remnants of the symmetry structure survive as trajectory-dependent invariants.

\section*{Acknowledgements}

The research of R. Azuaje is supported by the European Union and the Czech Ministry of Education under project CZ.02.01.01/00/22\_011/0008569 "Czech Technical University - International Postdoc Programme CROP".

A. M. Escobar Ruiz would like to thank the support from UAM research grant CBI-SA-391-26 PAPDI 2026.

\section*{AI Use Statement}

Language-model tools were used to assist with editing and symbolic checks. All mathematical statements, computations, and conclusions were independently carried out by the authors.

\bibliographystyle{unsrt}
\bibliography{references}

\end{document}